%% file: vis-stat.tex
\documentclass[11pt,twocolumn]{article}

\usepackage{times}

\usepackage[labelfont=bf]{caption}

\usepackage[top=0.5in,bottom=0.75in,left=0.5in,right=0.5in]{geometry}

\usepackage{url}

\usepackage{subfig}

\usepackage{graphicx}

\usepackage{amsmath}

\graphicspath{{figures/}{plots/}{.}}

\usepackage[numbers]{natbib}
\usepackage[bookmarksopen,bookmarksnumbered,colorlinks=true,allcolors=blue]{hyperref}

\usepackage{xcolor}

\usepackage[compact]{titlesec}

\usepackage{helvet}
\usepackage{sectsty}
\allsectionsfont{\sffamily}

\usepackage{mathptmx}   

\usepackage{flushend}

\usepackage{framed}

\usepackage[strict]{changepage}

\newenvironment{hquote}[1]{%
  \def\FrameCommand{%
    \hspace{1pt}%
    {\color{#1!35}\vrule width 2pt}%
    {\color{#1!35}\vrule width 4pt}%
    \colorbox{#1!10}%
  }%
  \MakeFramed{\advance\hsize-\width\FrameRestore}%
  \noindent\hspace{-4.55pt}
  \begin{adjustwidth}{}{7pt}%
  \vspace{2pt}\vspace{2pt}%
}
{%
  \vspace{2pt}\end{adjustwidth}\endMakeFramed%
}

\begin{document}


\title{\sffamily Visualization According to Statisticians: An Interview Study on the Role of Visualization for Inferential Statistics}

\author{\sffamily Eric Newburger and Niklas Elmqvist\\ 
\scriptsize\sffamily University of Maryland, College Park, MD, USA}

\date{\sffamily June 2023}

\maketitle

\begin{abstract}
    Statisticians are not only one of the earliest professional adopters of data visualization, but also some of its most prolific users. 
    Understanding how these professionals utilize visual representations in their analytic process may shed light on best practices for visual sensemaking.
    We present results from an interview study involving 18 professional statisticians (19.7 years average in the profession) on three aspects: (1) their use of visualization in their daily analytic work; (2) their mental models of inferential statistical processes; and (3) their design recommendations for how to best represent statistical inferences.
    Interview sessions consisted of discussing inferential statistics, eliciting participant sketches of suitable visual designs, and finally, a design intervention with our proposed visual designs.
    We analyzed interview transcripts using thematic analysis and open coding, deriving thematic codes on statistical mindset, analytic process, and analytic toolkit. 
    The key findings for each aspect are as follows: (1) statisticians make extensive use of visualization during all phases of their work (and not just when reporting results); (2) their mental models of inferential methods tend to be mostly visually based; and (3) many statisticians abhor dichotomous thinking. 
    The latter suggests that a multi-faceted visual display of inferential statistics that includes a visual indicator of analytically important effect sizes may help to balance the attributed epistemic power of traditional statistical testing with an awareness of the uncertainty of sensemaking.
\end{abstract}


\begin{figure*}[tbh] 
  \centering
  \includegraphics[width=\textwidth]{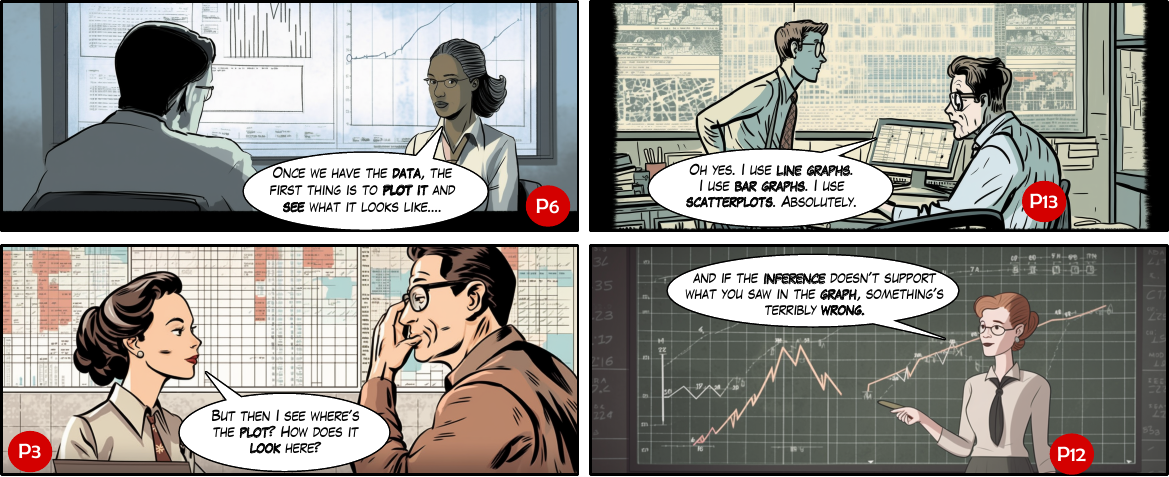}
  \caption{\textbf{Visualization as a bridge.}
  Understanding how professional statisticians use and think about visualization may help designing effective visualizations to support sensemaking for everyone. 
  Note that these characters bear no likeness to the original participants and their quotes have been slightly edited for brevity.
  (Images by MidJourney v5.)
  }
  \label{fig:teaser}
\end{figure*}

\input{content/01-intro}

\input{content/02-background.tex}

\input{content/03-method.tex}

\input{content/04-results.tex}

\input{content/05-discussion.tex}

\input{content/06-conclusion.tex}

\section*{Acknowledgments}

We thank the anonymous participants for their time, effort, and willingness to share their knowledge and wisdom. We also thank Jennifer Cheeseman Newburger for her parallel coding work.

\bibliographystyle{plainnat}
\bibliography{vis-stat}

\end{document}

%% file: content/01-intro.tex
\section{Introduction}

Statistics was one of the early adopters of data visualization, and graphical methods in themselves are a valid form of inference~\cite{Cleveland1993}.
Even today, statisticians remain some of the more prolific users of data visualization, with certain statistical tests routinely involving visual inspection and graphical inference.
As a case in point, John W.\ Tukey's 1977 book on \textit{exploratory data analysis} (EDA)~\cite{Tukey1977} established the field of \textit{visual statistics}~\cite{Buja2009}, where interactive visual representations are used to inform and generate hypotheses, or even confirm them. 
While workflows differ between each practicing statistician, it is clear that most have an intimate and working knowledge of visualization for sensemaking.

Interestingly, visualization is also commonly described as a key enabling technology for helping people understand and make decisions based on data~\cite{DBLP:journals/tvcg/GrammelTS10, DBLP:journals/tvcg/HuronJC14, DBLP:journals/tvcg/PousmanSM07}.
Unlike arcane statistical tests and mathematical formalism, interactive visual representations straightforwardly invite users to overview, filter, and drill into~\cite{DBLP:conf/vl/Shneiderman96} data with little prerequisite knowledge. 
Given appropriate prompting and visual representations, even laypeople can manually perform statistical tests such as comparing averages in time-series data~\cite{Correll2012, Albers2014}, fit trend lines to point clouds~\cite{conf/chi/CorrellH17}, and make mean value judgments in multi-class scatterplots~\cite{Gleicher2013}.
The prevalence of visualization for both novices and experts alike suggest that \textbf{visualization could become a bridge for giving people access to advanced statistical workflows}.

In this paper, we perform an interview study with professional statisticians to understand their practices for analyzing and making decisions using visualization.
Our research team is well placed to do this work: the first author is a former practicing statistician with a career spanning two decades in the U.S.\ Census Bureau.
Using this author's professional network, we recruit a total of 18 statisticians with a combined 350 years of experience (average 19.7 years).
We focus on three questions:

\begin{itemize}
    \item[RQ1] How do statisticians use visualization in their daily analytic work?
    \item[RQ2] What mental models of inferential statistics do statisticians have?
    \item[RQ3] How to design a representation for statistical inference that builds on current practices of professional statisticians?
\end{itemize}

Our interview study was conducted as a semi-structured interview over Zoom videoconference.
Each session involved three phases: (1) statistical practice; (2) graphical elicitation of their internal understanding of inferential statistics; and (3) review of a design probe~\cite{DBLP:journals/interactions/GaverDP99}: a prototype visual representations for statistical inference.
All sessions were professionally transcribed and the first author used their statistical expertise and experience to code the transcripts using an open-coding approach~\cite{DBLP:books/el/LFH2017}.
We then derived our findings using thematic analysis.

At a high level, we found that visualization tends to be a key activity in most statistician's daily workflow, and not just during presentation.
Furthermore, our participants mostly reported mental models that are visually based.
Finally, we found that our participants tend to abhor dichotomous thinking and distrust insights lacking multiple evidence.

%% file: content/02-background.tex
\section{Background}

Here we review the key background research on statistical inference, visual statistics, and visualization.

\subsection{Statistical Inference}

Inferential statistics involves using statistical methods to make inferences about a population based on a sample of data~\cite{casella2001statistical}.
The field can be traced back to the work of Ronald Fisher, who is considered one of the founding fathers of modern statistics~\cite{Hald1998, Fisher1922}.

The traditional approach to statistical inference is \textit{confirmatory data analysis} (CDA)~\cite{Lehmann2005, Schervish1995}, 
which typically involves specifying a hypothesis, and using a statistical test to determine whether a counter (null) hypothesis might explain the data.
CDA is often used in experimental research and is a critical component of modern science. 

\textit{Exploratory data analysis} (EDA), on the other hand, first coined by John Tukey in the 1970s~\cite{Tukey1977}, is an approach to analyzing data that involves exploring and summarizing the data to gain insights and identify patterns.
Tukey argued that EDA was an essential first step in data analysis that is necessary for understanding the data before applying confirmatory techniques.
His work became part of the foundation for visualization, where interactive visual representations are used to inform, generate, or even confirm hypotheses.
Most statisticians agree that both EDA and CDA have important roles to play in data analysis.

\subsection{Visual Statistics}

There are many routine tasks for which a practicing statistician will turn to graphical methods~\cite{Cleveland1993}, including model selection (using residual plots or Q-Q plots to verify assumptions), outlier detection (using scatterplots or histograms), and quality control (plotting the data).

There has been a growing interest in the visualization community in how graphical representations of data can facilitate higher-order tasks, such as making inferences.
To address this issue, Buja et al.~\cite{Buja2009} proposed frameworks for \textit{visual statistics}, which rely on visual representations to serve as a test statistic while human cognition functions as the statistical test.
They demonstrated their approach using what amounts to a ``Rohrschach test'' of random data in a lineup of small multiples, where only one of the multiples utilized real data.

In subsequent research, Wickham et al.~\cite{Wickham2010} adapted this concept for the visualization community and discussed how it can be applied to common visualizations to reveal new insights while minimizing false positives.
Beecham et al.~\cite{Beecham2017} implemented the lineup protocol for graphical inference in geographic clustering visualizations, while Correll et al.~\cite{Correll2019} used it to examine the effectiveness of common distribution graphics in displaying gaps or outliers.

\subsection{Everyone a Statistician}

Novices often struggle with choosing methods, understanding assumptions, interpreting results, and implementing tests when they are in need of statistics to analyze and understand the data they deal with in their everyday lives~\cite{Mustafa1996}.
Interestingly, a visual approach to statistics can help even these novice users perform advanced statistical tests~\cite{DBLP:journals/tvcg/GrammelTS10, DBLP:journals/tvcg/HuronJC14, DBLP:journals/tvcg/PousmanSM07}.
As a case in point, recent work has shown how even novice users can compare averages in time-series data~\cite{Correll2012, Albers2014}, fit trend lines to point clouds~\cite{conf/chi/CorrellH17}, and make mean value judgments in multi-class scatterplots~\cite{Gleicher2013} without specialized training or knowledge.

The theoretical basis of our work is grounded in the observation that visualization represents common ground between novices and experts alike, and thus that visualization can become a bridge for giving people access to advanced statistical workflows.
However, to achieve this, we must first understand the visualization practices of expert statisticians.
Below we will describe our approach to achieve this goal.

%% file: content/03-method.tex
\section{Method}

We conducted a qualitative study through semi-structured interviews with professional statisticians.
Our questioning focused on participants' relationship with visualization and their understanding of statistical inference with an emphasis on frequentist, parametric statistical tests (even if many expressed their understanding of non-parametric and/or Bayesian approaches as well).
Data were coded according to a composite scheme that mixed a priori and emergent codes.

\subsection{Positionality Statement}
\label{sec:positionality}

Given the qualitative nature of our evaluation study, the positionality of the authors may have had an impact on our reporting and interpretation of our findings.
The first author is a recent Ph.D. in information science and a former statistician with the U.S.\ Federal government, and the second is a faculty member in information and computer science.

We have implemented the following strategies to mitigate the impact of these potential biases:

\begin{itemize}

    \item During \textbf{participant recruitment}, we drew from a mix of analytic communities to ensure a diversity of analytic viewpoints, but also to reach statisticians with experiences separate from ours.

    \item While \textbf{creating the collection instrument}, we mixed three different collection modes (open-ended questions, graphic elicitation, observations on designs probes).  
    This diversity of data types gives the data collection resilience against potential biases.

    \item In \textbf{conducting interviews}, we endeavored to create an environment which would provide participants the comfort of feeling they were having a conversation with an interested and supportive colleague; to both create a space in which they could openly discuss the most technical aspects of their work without fear of alienating their listener, and safely share private thoughts about their work.
    There is evidence this environment succeeded, as more than one participant expressed their relief that the study's anonymity precautions would ensure none of their employers would know what they had been saying. 

    \item During \textbf{coding of the findings}, we used a combination of a priori and emergent codes. A priori coding required us  to take a disciplined approach to some of the results, with the potential to directly falsify our initial hypotheses. Emergent coding encouraged us to think beyond these initial hypotheses.  

\end{itemize}


\begin{table}[tbh]
    \centering
    \includegraphics[width=\linewidth]{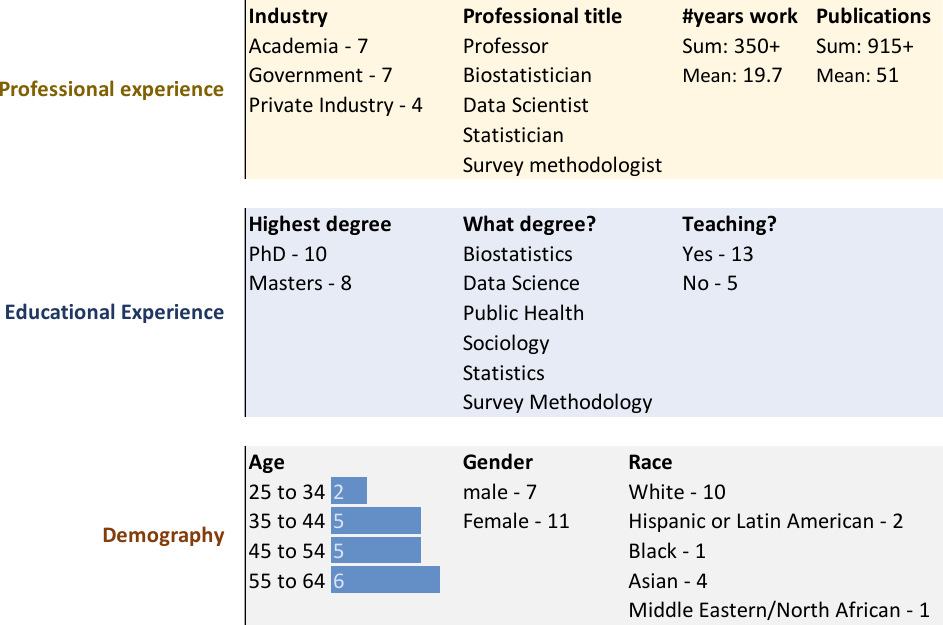}
    \caption{\textbf{Interview study summary.} 
    Summary of participants and their demographics in our interview study.
    }
    \label{tab:study-summary}
\end{table}

\vspace{-5mm}
\subsection{Participants}

We recruited 18 paid participants via direct email request.
Participants were offered \$25 compensation in the form of a gift card.
Two thirds of participants were selected from among members of the American Statistical Association, with guidance from the organization's leadership.
With one exception, these participants had no prior professional contact with the study authors.  
The remaining one third of participants were reached through the authors' professional networks.

Our recruitment efforts explicitly targeted statisticians working in three broad industries: government statistical agencies, academia, and private industry.
We used the following screening criteria: at least 18 years of age; at least one relevant degree (undergraduate or graduate);
5 or more years of experience as a professional statistician after the completion of their education; and job duties that included statistical inference, such as statistical tests or confidence intervals.


\begin{table}[tbh]
    \centering
    \includegraphics[width=\linewidth]{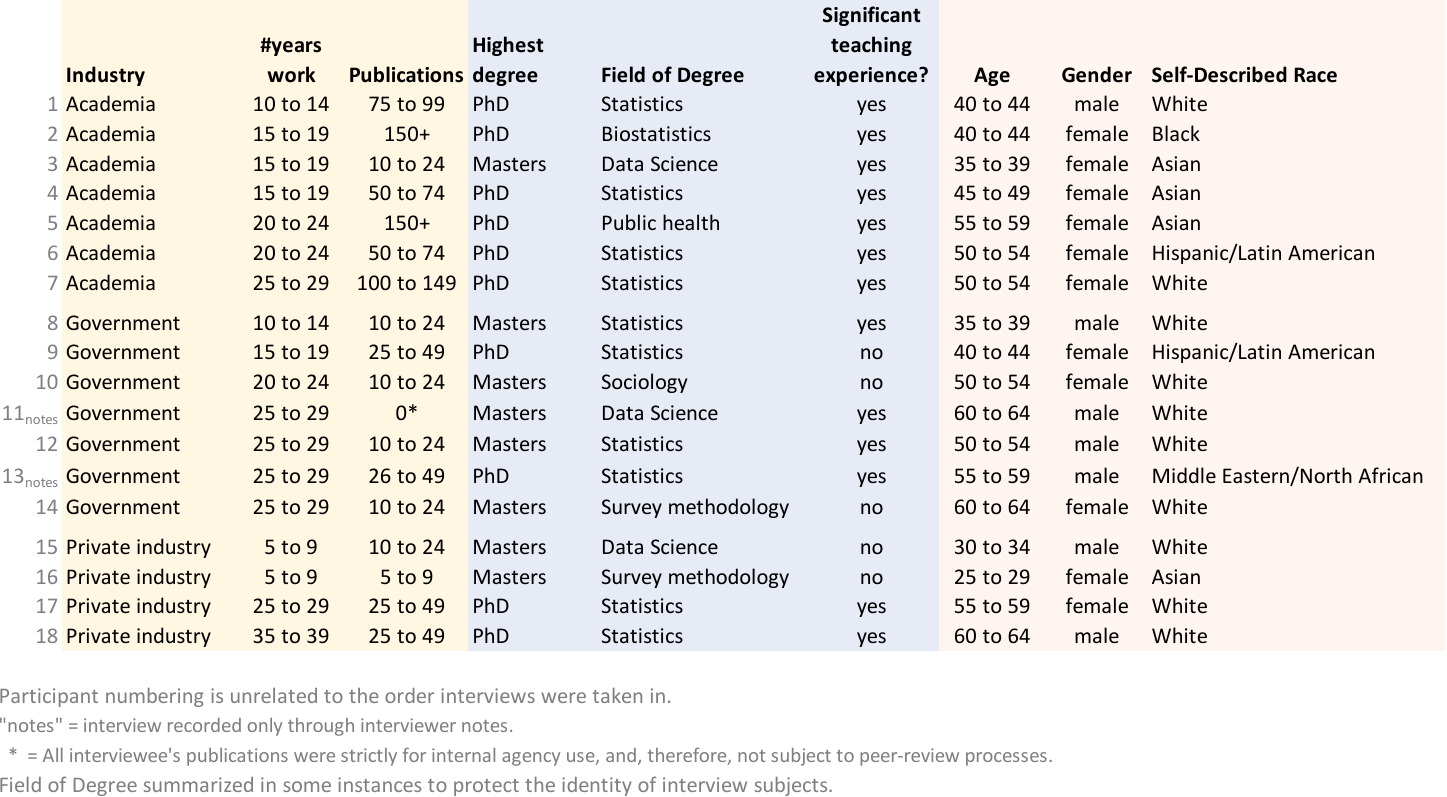}
    \caption{\textbf{Participant overview.} 
    Demographics, experience, and training for the 18 participants in our interview study.
    }
    \label{tab:participants}
\end{table}

Table~\ref{tab:study-summary} summarizes the study and Table~\ref{tab:participants} gives an overview of the participants.
As can be seen from the tables, partipants varied in their professional experience, educational experience, and demographics.



\subsection{Data Collection}

All interviews were conducted via live videoconferencing on Zoom using a laptop or desktop computer.
Sessions were both video and audio recorded.
Video allowed for screen-captures of sketches drawn by participants.
Audio recordings were submitted to a transcription service (\url{http://rev.com}) to provide accurate text for coding.
The researcher took notes during the sessions, including sketches of participant-described visualizations, which the researcher then showed to the participants via the video feed for their approval of the sketches.
The interviewer also took notes immediately after each session to record general impressions, and recall details not captured in the moment. 

There were a few exceptions to the interview protocol.
The first participant used their smartphone for the videoconferencing, which resulted in lower fidelity transmission of graphics for the third section of the interview.
Two of the 18 interviews failed to record, and, therefore, only the researcher’s notes captured the outcomes of these two sessions.
One interview skipped the graphic elicitation section of the interview, while completing the other two sections in full.

\subsection{Interview Script}

Interviews followed a script (see supplemental material).
Questions in bold were asked word for word of all participants, optional follow-up questions appear indented in the table.   
Prior to data collection, the researchers conducted two practice interviews using early versions of the interview script, and a sketch-version of a strawman graphic (see below).
Scripts and graphic were refined based upon feedback during these practice runs, and a second strawman graphic was added.  

\subsection{Design Probes: Strawman Graphics}

One phase of our evaluation involved the use of \textit{design probes}~\cite{DBLP:journals/interactions/GaverDP99, DBLP:conf/chi/WallaceMWO13} to elicit feedback from our expert participants.
In this case, our design probes were visual representations designed for supporting graphical inference by experts.
We call these probes ``strawman graphics'' because we hope that participants will have constructive feedback on each. 

During the course of our study, we ended up developing three separate strawman graphics (labeled \#1, \#2, and \#3) based on participant feedback and suggestions collected from the interviews.
The first was designed prior to the first session, the second during the pilot interviews, and the third came out of the first four interviews.
We report on the design iterations of each graphic in the results section.

\subsection{Procedure}


\paragraph{Pre-interview.}

Scheduling was conducted over email.
Participants were asked to fill out a consent form before their interview.
This included a small number of demographic and qualifying questions.

\paragraph{Interview: Preliminaries.}

At the start of each session before recording began, participants were given the opportunity to ask any questions they had.
With these preliminaries cleared, recording began.

\paragraph{Interview: I - Analytic Process (RQ1).}

We first asked participants to recall the steps of their analytic process.  
This framed subsequent discussions as focused on their day-today work process.
Probes into these processes sought to capture how visualization played a part (or not) in their workflow.
Additional probes sought to uncover the statisticians’ tacit understanding of their processes. 

\paragraph{Interview: II - Graphic Elicitation (RQ2).}

Here we used any references to inferential statistics from the prior interview section as a bridge to turn the conversation toward participants’ understanding of inferential statistics in general, and Student t-tests~\cite{Student1908} in particular.
Participants were asked whether they had a picture in their heads of what a two-mean t-test looked like.
They were then asked to draw a picture of that image and walk the interviewer through their sketch.  

\paragraph{Interview: III - Strawman Graphics (RQ3).}

The interviewer then presented participants with the design probes: strawman graphics of a t-test constructed by the research team.
Participants were walked through each version (2 versions for the first three interviews, with strawman \#3 added to the rest based upon the early results).  
They were first asked to use the graphics to answer whether the displayed example met a selected significance level ($p = \{ 0.1, 0.05, 0.01, 0.001 \}$ for strawman \#1 and \#2 and only $p = .05$ for strawman \#3).
The same underlying data were displayed with all three graphics, so that differences in response would reflect the graphic rather than a difference in data samples.
After capturing participant performance in using each graphic, ‘correct’ answers were provided before moving onto the next graphic.
Afterward, participants answered a series of questions designed to probe for their understanding or initial misunderstanding, what was missing or wrong about the graphics, or other design recommendations.   

\paragraph{Interview: Closing.}

The interviews ended in two phases.
Participants were first thanked, and asked on camera whether they had additional questions or reactions to share.
They were then given the opportunity to share unrecorded feedback. 

\subsection{Coding Process}

We used thematic coding methodology derived from grounded theory~\cite{DBLP:books/el/LFH2017} (although we did not use the full grounded theory machinery).
The three parts of the interview were coded by separate processes, with some overlap.
We used a combination of open and closed coding~\cite{DBLP:books/el/LFH2017}; the details of the coding scheme are given in the results section.
Here is the rough procedure we followed:
We first reviewed our notes, looking for themes.
Based upon the themes, we created the coding schema (Table~\ref{tab:coding-scheme}) combining a priori codes with open spaces for capturing emergent themes. Tables 3 and 4 denote open codes with '(list)'; other codes are a priori.
We then coded the transcripts, first by deriving primary codes and refining them into detailed thematic codes.
Finally, we grouped all codes into broad themes.

To check the validity of the coding process, we engaged a second coder external to the research team.
Given the nature of the data, this coder had to be a statistician of similar experience to the researchers and participants.
The external coder was provided with a 10\% random sample of texts (with codes stripped), along with the final state of the codebook.
After calculating the inter-coder reliability, we conferred with the external coder to update the codebook appropriately.

%% file: content/04-results.tex
\begin{table}[tbh]
    \centering
    \includegraphics[width=\linewidth]{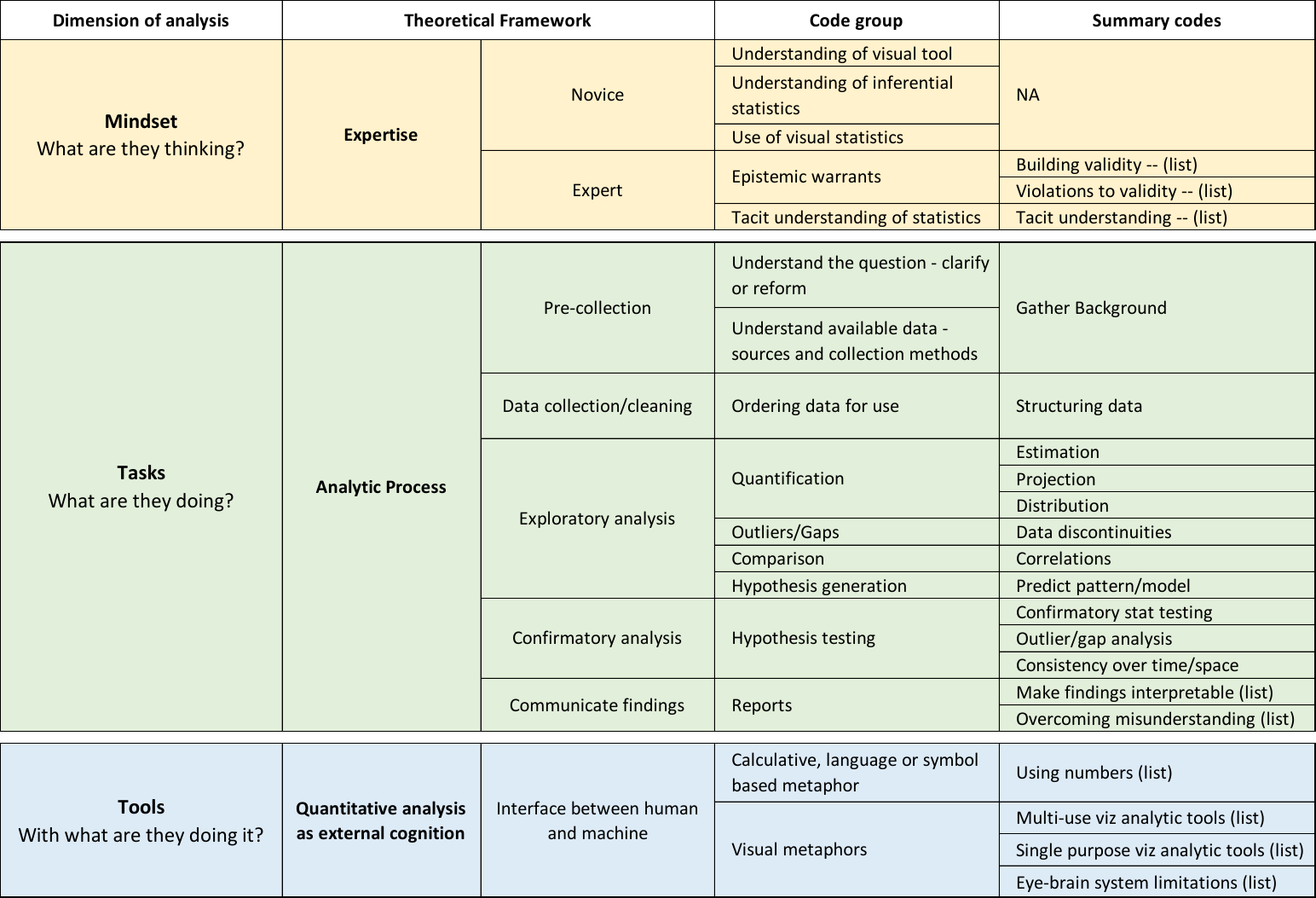}
    \caption{\textbf{Coding scheme.}
    Overview of the coding scheme developed for the interview transcripts.
    }
    \label{tab:coding-scheme}
    \vspace{-5mm}
\end{table}

\section{Results}

We transcribed and coded interviews for all participants.
Here we describe the coding scheme and the design probes.
Then we give an overview of the coding results and the inter-rater reliability.
Finally we report our detailed findings for each of the themes.

Note that this treatment only gives a high-level overview of our findings; the supplemental material contains all of the detailed results (including additional tables and charts).

\subsection{Coding Scheme}

We applied three parallel coding schema to the main body of the interviews based upon three analytical perspectives: expertise, tasks, and tools.
Each perspective answers a different fundamental question about participant expectations from inferential statistical methods, potential visualizations of inferential statistics, and visualizations in general.

\begin{itemize}
    \item \textbf{Expertise perspective:} The explicit as well as tacit knowledge that professional statisticians have about their field.
    
    \item \textbf{Tasks perspective:} Analysis steps that statisticians pursue.
    
    \item \textbf{Tools perspective:} The tools statisticians employ in their work.
\end{itemize}

\subsubsection{Coding Process}

The three perspectives constitute what amount to three different dimensions of analysis, orthogonal to one another, any or all of which might be indicated by a single participant statement. 
For example, P7 described part of their analytic process as:

\begin{quote}
    \textit{And then once we got the data, [...] half the time is understanding the data [...]
    using descriptive statistics, a graphical illustration, [to see] missing data pattern or some unusual outliers.}
\end{quote}

This response generated the following summary:

\begin{itemize}
\item \textbf{Expertise} -- Preparing and understanding data is a major task.
\item \textbf{Tasks} -- Structure, background, distribution, estimates, etc.
\item \textbf{Tools} -- Multitool: descriptive graphics.
\end{itemize}

We then reviewed the summary to produce these primary codes:

\begin{itemize} 
\item \textbf{Expertise} -- \textsc{Data prep is much of stat work} (Tacit).
\item \textbf{Tasks} -- \textsc{Structure, Background, Distribution, Estimates, Discontinuities}.
\item \textbf{Tools} -- \textsc{Descriptive Graphics}, a broad class of visualizations.
Had a specific tool (such as \textsc{Histogram} or \textsc{Scatterplot}) been mentioned, these would also have been coded.
\end{itemize}

We further grouped primary codes from the expertise perspective into a detailed thematic code: \textsc{Reality is the authority}, which we further group under the broad theme \textsc{Warrant}.

\subsubsection{Expertise Perspective}

The \textbf{Expertise} perspective contrasts the mindset of experts with that of novices.
During interviews, the perspective can be summarized by the question, ``What is the participant thinking during their interaction with inferential statistics?''
The answers come under two broad groupings:

The first group of codes captured the \textsc{Epistemic Warrants} experts attribute to different inferential methods, that is, the degree to which a particular analytic method provides support for or against some claim.
Two emergent codes made up this group.

\begin{itemize}
    \item \textsc{Build Validity} captures those elements of analysis that shore up a reasoned line of evidence.
    
    \item \textsc{Violations to Validity} captures those elements of analysis that tend to undercut evidence.
\end{itemize}

The second code group includes only a single emergent code: \textsc{Tacit} captures the knowledge that experienced statisticians have, through work experience, come to believe about their method.

\subsubsection{Tasks Perspective}

The \textbf{Tasks} perspective is composed primarily of a priori codes.
It emerged out of initial reviews of the interview results, and presumes that there are common steps in the analytic process all statisticians pursue, regardless of the sub-field in which they work.  
Identifying the core tasks of analysis as perceived by experienced statisticians provides a context in which to place the kinds of inferential statistics tools this research effort hopes to design, and may suggest features the design should incorporate.
More broadly, the Tasks perspective sets the context for understanding the other two dimensions of analysis. 

The initial review of interviews indicated five analytic phases within Tasks (Table~\ref{tab:coding-scheme}).  
As the focus of this research is inferential statistics, a later stage part of the analytic process, we sacrificed some detail during coding of the earliest work phase. 
This meant that we consolidated the code groups of understanding the research questions and understanding data sources into a single code, \textsc{Gather Background}.

While we initially expected this perspective to only consist of a priori codes, we found during early review that the \textsc{Communicate Findings} codes (labeled \textsc{Myth} and \textsc{Comm}) often included details of what needed to be communicated that gave further insight into the mindset of participants.
Thus, these were recoded as emergent codes, with details captured and summarized for thematic coding using the Expertise (mindset) thematic codes list.

\subsubsection{Tools Perspective}

The \textbf{Tools} perspective captures the specific interfacing-tools statisticians choose to employ during their work.
Broadly speaking, these tools can either be \textsc{Visual Metaphors} (data visualization) or \textsc{Symbolic Tools} (such as code or written equations).

The \textsc{Visual Metaphor} code group includes three summary codes:

\begin{itemize}
    \item \textsc{Broad-focus visualizations} that may have many uses (such as histograms, from which an analyst might discern the mean, median, variance, skewness, range, location and number of modes, gaps in the distribution, outliers, and shape);
    \item \textsc{Narrow-focus visualizations} with a single use (such as QQ-norm plots, which are used almost exclusively to test the normality of a distribution); and
    \item \textsc{Limits}: observations on the limitations of visualization and the human visual system offered by participants.
\end{itemize}

Since visualization is the focus of this research, we captured the non-visual \textsc{Symbolic Tools} code group with only one summary code and then listed each tool by name.

\subsubsection{Graphic Elicitation Coding}

Participant graphics were screen-captured along with participant descriptions of the images.
We labeled these by type and then grouped them according to their analytic focus.  

\begin{table}[tbh]
    \centering
    \includegraphics[width=\linewidth]{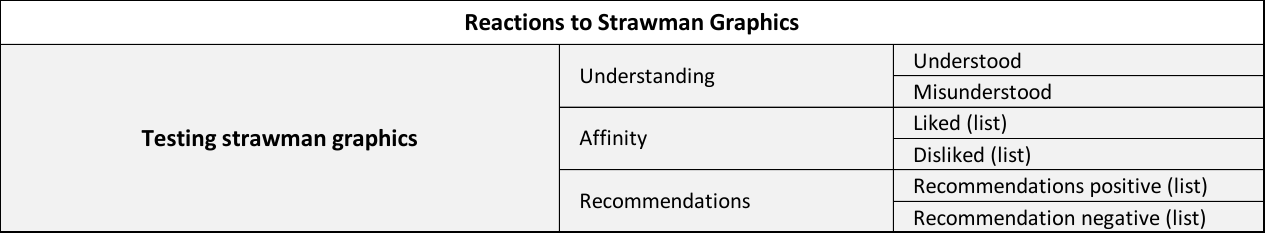}
    \caption{\textbf{Strawman coding scheme.}
    Overview of the coding scheme developed for the strawman graphics.
    }
    \label{tab:strawman-coding-scheme}
\end{table}

\subsubsection{Strawman Graphics Coding}

The strawman graphics section was coded using a combination of a priori and emergent codes (Table~\ref{tab:strawman-coding-scheme}).
These codes captured as follows:

\begin{enumerate}
    \item \textbf{Participant Understanding}: Whether the participant correctly assessed the significance at the given alpha value with the graphic;
    
    \item \textbf{Participant Affinity for the Strawman Graphic}:  Whether participants found the strawman graphic useful, informative, or otherwise saw it in a positive or negative light; and

    \item \textbf{Participant Design Recommendations}: Any directly stated or implied design recommendations to improve the graphic.
\end{enumerate}

The intention of this phase was to generate informative conversation about, and observations on, potential elements for graphic inference tools.
By asking participants to actually use a graphic tool to perform a statistical inference, we hoped to generate deeper insights than merely asking them to comment on a novel graphic.
This worked to such an extent that, in addition to revealing several design recommendations, this section of the interviews generated responses that were informative of expert statisticians understanding of inferential statistics in general.

\subsection{Strawman Graphic: Design Iterations}

We produced a total of three versions of the strawman graphics used as design probes during the interview study. 

\begin{figure}
    \centering
    \subfloat[Strawman \#1]{
        \includegraphics[width=0.45\linewidth]{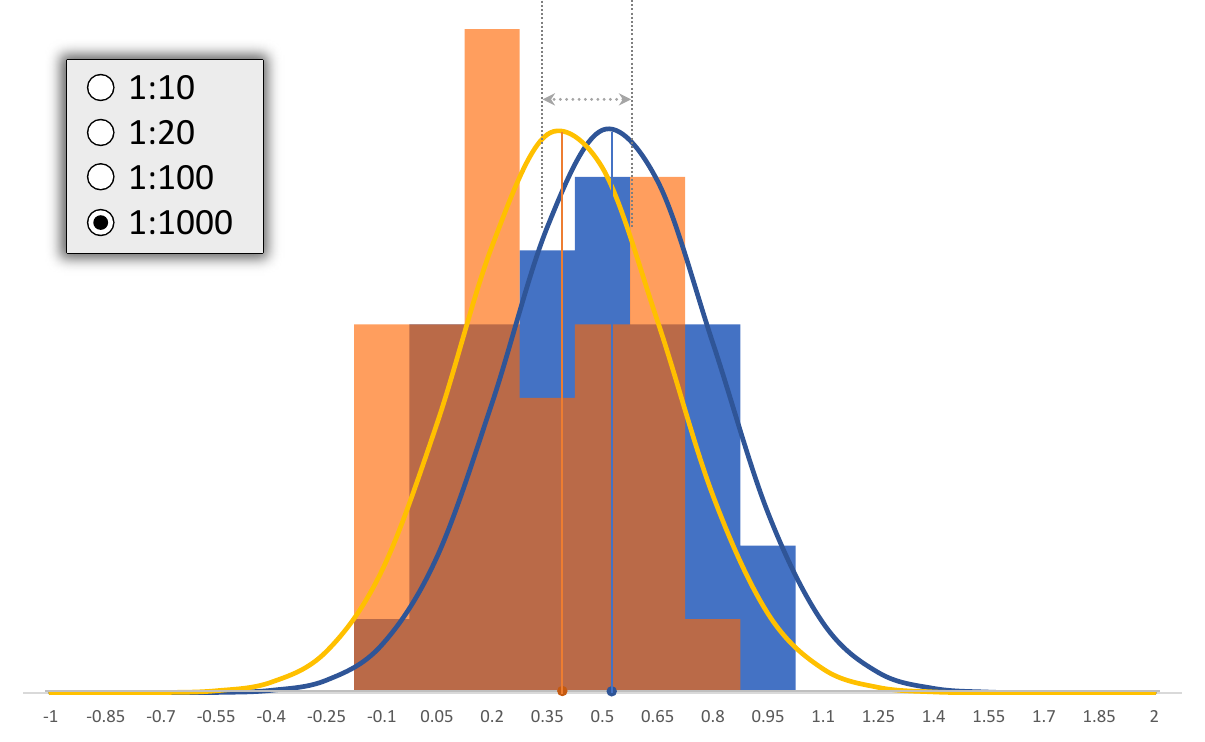}
        \label{fig:strawman1}
    }
    \subfloat[Strawman \#2]{
        \includegraphics[width=0.45\linewidth]{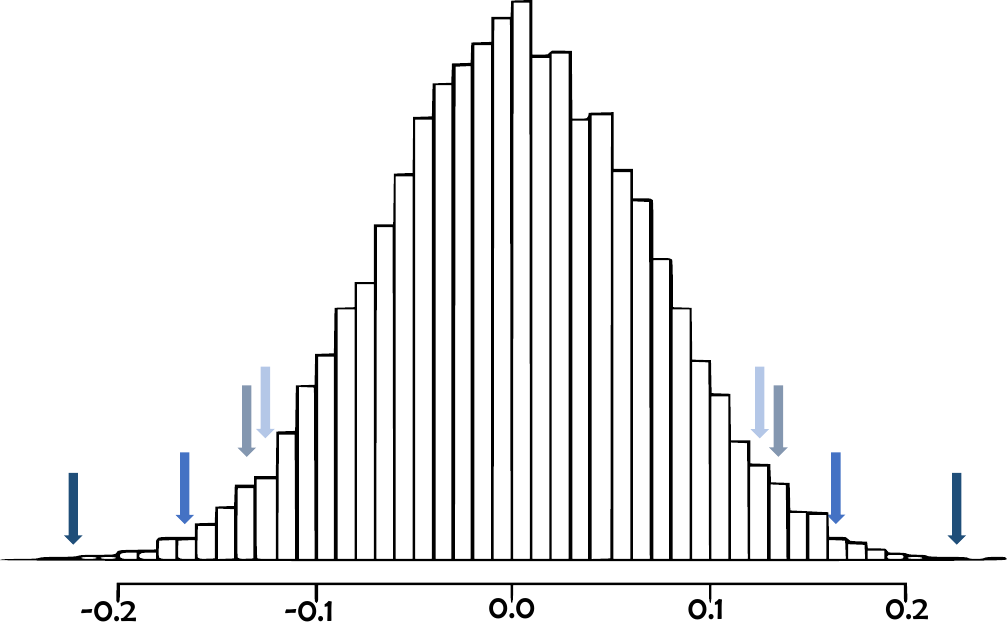}
        \label{fig:strawman2}
    }\\
    \subfloat[Strawman \#3]{
        \includegraphics[width=\linewidth]{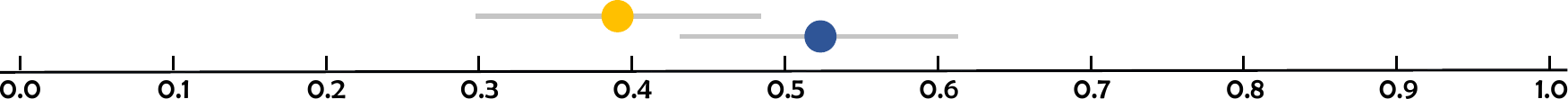}
        \label{fig:strawman3}
    }
    \caption{\textbf{Strawman graphics.}
    The graphics we used as design probes.
    }
    \label{fig:strawman-graphics}
\end{figure}

\paragraph{Strawman \#1.}

This graphic illustrates the overlap of two sample distributions---the blue and orange samples---represented both as a pair of histograms and as a pair of normal curves that had been fit to the samples (Figure~\ref{fig:strawman1}).
The goal for users was to determine whether the two samples were similar enough in their means that they were likely drawn from a common population, or whether they were so unlikely to have been drawn from a single population that they probably represented sub-populations.
The graphic included an aid for users to make this determination in the form of a \textit{difference ruler}: a pair of grey lines connected by a double-headed arrow signifying the amount of separation two sample means needed to show to represent a statistically significant difference for a given confidence level (alphas corresponding to a $p$-value of .1, .05, .01. or .001).
We drew on our own work on fitting bell curves~\cite{Newburger2023} in designing overlapping histograms to represent a statistical test of two samples.

\paragraph{Strawman \#2.}

This iteration (Figure~\ref{fig:strawman2}) was suggested by a participant during one of the pre-interviews.
It attempts to represent the distribution of the test statistic for the two sample means, with arrows to indicate how far out on the distribution a test statistic needed to fall to indicate statistical significance at a given alpha level.
The graphic was created using a simulation process, in which 10,000 pairs of samples were drawn from a normally distributed population with a mean set as the mean of the two original samples (orange and blue and a variability (standard deviation) as the joint variability of the two original samples.
For each pair of samples, a difference of means was calculated and these mean differences were plotted in a 50-bin histogram to yield a fairly smooth distribution.
We added arrows to indicate the bar in which the mean-difference resided representing the minimum difference required for statistical significance at a given confidence level. 

\paragraph{Strawman \#3.}

Three of the first four participants indicated that some version of overlapping confidence intervals was their internal model of a t-test.
Inspired by these interviews, the third iteration was designed as a pair of overlapping 95\% confidence intervals (Figure~\ref{fig:strawman3}).
Therefore, we included strawman \#3 in all subsequent interviews. 

\begin{table}[htb]
    \centering
    \includegraphics[width=\linewidth]{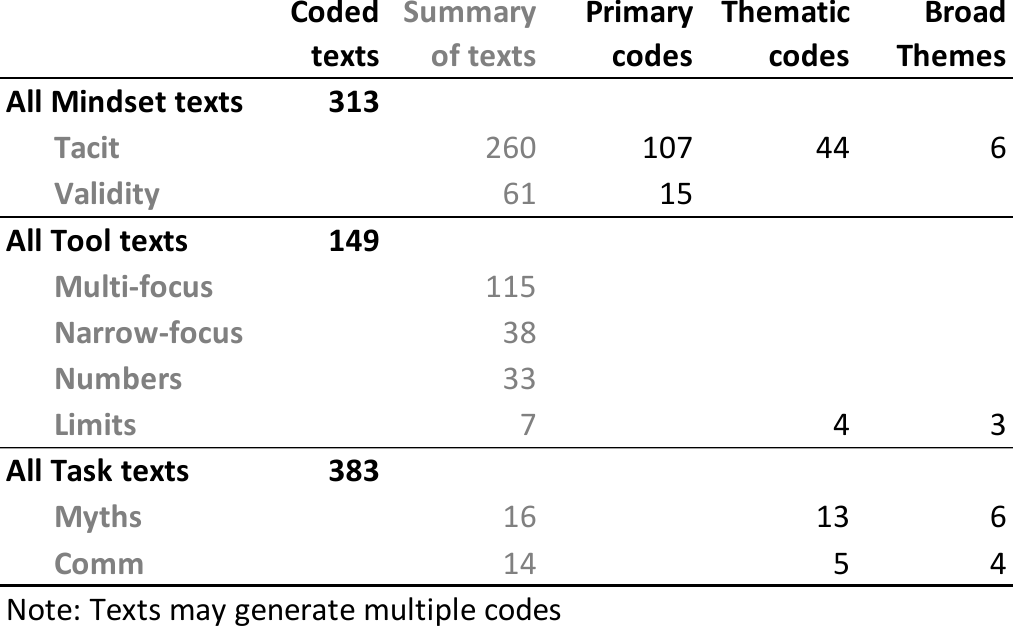}
    \caption{\textbf{Coding overview.}
    Summary of codes in interview transcripts.
    }
    \label{tab:coding-overview}
\end{table}

\subsection{Overview} 

Transcripts from 16 participants were assigned a total of 845 codes.
Table~\ref{tab:coding-overview} gives an coding overview for each of the three perspectives.
In all, 313 texts were assigned codes associated with the Expertise perspective.
Tool codes amounted to 149 for mentions of specific tools, including 115 listing broad-focus tools, 38 narrow-focus, and 33 symbolic ones (numbers).
Finally, there were 383 occurrences of Task codes.
This likely reflects the structure of the interviews, which was focused on the analytic process.
Some mentions of specific tools and analytic tasks were also collected from researcher notes. 

\subsection{Intercoder Reliability}

Coding was tested for intercoder reliability via a 10\% sample of transcripts.
Since coding was done on individual texts, the context of the full interview---such as participant comments just prior to and following each text in sample---are missing.
This may have reduced reliability. 

The validity code agreement between coders was 71\%; Cohen’s Kappa was found to be .588 (moderate agreement).
\textsc{Tacit} coding was also tested for intercoder reliability via a 10\% sample at the broad themes and detailed thematic code levels.
Among broad codes, there was 67.7\% agreement, with a Cohen’s Kappa of .611 (substantial agreement).
Agreement between coders for thematic codes was 51.7\%, with Cohen’s Kappa at .495 (moderate agreement).   

\begin{table}[htb]
    \centering
    \includegraphics[width=0.8\linewidth]{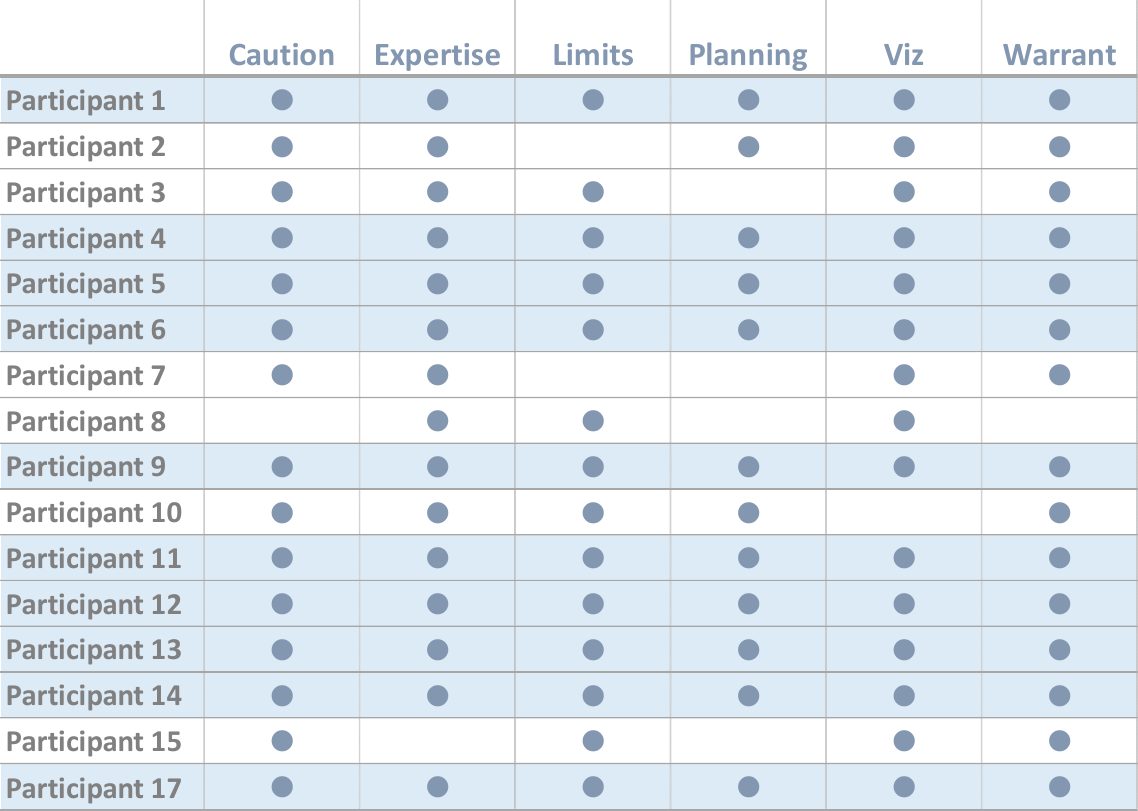}
    \caption{\textbf{Broad themes.}
    Participant coding for the six broad themes.
    }
    \label{tab:broad-themes}
\end{table}

\subsection{Broad Themes}

The six broad themes have wide support from participant interviews (Table~\ref{tab:broad-themes}).
A majority of participants made comments coded into all six, and no participant made comments coded into fewer than three. 

\begin{itemize}
    \item \textsc{Caution.}
    This theme expressed the multifaceted concerns statisticians have in pursuing their work, and their approaches to addressing those concerns.
    It included thematic codes such as \textsc{People act upon our results}, an admonition to remember that people trust statistical work, take action based upon it, and, therefore, it is a statistical professional’s responsibility to put in whatever time and effort is required to always provide the best possible advice.
    The group also includes, \textsc{Distrust findings}, which captured the many ways participants remind themselves to always check and recheck their work, since statistical analysis is a fundamentally complex endeavor which allows for many points of failure.
    The \textsc{Caution} theme also touched upon statisticians’ relationship to inferential statistical methods, with, \textsc{P-values encourage bad thinking}, referring to concerns in the statistical community about the tendency for statistical testing to foment dichotomous thinking about complex realities, and, \textsc{Stat tests required for publication}, expressing participants' belief that whatever risks parametric statistical tests entail, they are nonetheless required for acceptance within many scientific communities, and thus must be used as best they can.

    \item \textsc{Expertise.}
    This group applies to the several aspects of acquired analytic understanding which, as a body, represent a divide between statistical professionals and people outside the field.
    It includes codes such as, \textsc{Analysis takes a statistician}, capturing participants' expressions of their belief that people outside the statistical field typically misunderstand at least some aspects of quantitative work.
    It also included, \textsc{Stat testing is hard for statisticians, too}, which captured participants’ expressions that statistical inference is a subject so complex that they don’t trust their own knowledge without the use of references. 

    \item \textsc{Limits.}
    This theme captured several observations from participants describing ways that details of a data collection can limit the range of statistical tools available to apply, but also the ways in which the choice of statistical methods can limit the scope of analytic research.
    These codes are not specifics about the various limitations discussed, but, rather, the awareness among participants that quantitative work entails limitations.
    Example codes include \textsc{Sample size important}, expressing the multiple dependencies between sample size and the validity of inferences made about sampled populations, and, \textsc{Stat methods define scope}, capturing expressions of how the tools of statistics define the kinds of questions statistical research can address.  

    \item \textsc{Planning.}
    These codes capture the importance of planning in quantitative work: its utility, costs, and pitfalls.
    For example, the code, \textsc{Predict to escape rationalization,} captures the participants’ understanding that post hoc rationalization 
    is a constant temptation during analytic work which threatens results validity, and that the way to avoid this through planning ahead; they plan the analyses they will run, the test statistics they will accept, etc.
    \textsc{Plan defines scope}, captures the participants’ awareness of how the planning process, while vital to the work, once entered, limits possible discoveries the work may yield.  

    \item \textsc{Viz.}
    This theme captures participants’ understanding of data visualization as a tool in their analytic work.

    \item \textsc{Warrant.} 
    These codes capture participants’ sometimes contradictory understandings of what elements within, or conditions are required by, their quantitative work to support statements about the world.
    These codes can be broad, such as, \textsc{Reality is the authority}, expressing participants’ focus on always connecting their computations as directly as possible to the subject of their study, or checking results against expected values extracted from ``facts on the ground'' sources, such as news reports.
    Some are more specific, such as, \textsc{Effect size $\geq$ p value}, which expressed the common feeling among participants that statistical significance was less important, or at least no more important, than the practical significance in their results.
    For example, a trial on a cholesterol drug with a large enough sample size might show a statistically significant reduction in blood cholesterol levels, but that reduction could still be so small as to have no expected effect on clinical outcomes for patients.    

\end{itemize}

\begin{table}[htb]
    \centering
    \includegraphics[width=\linewidth]{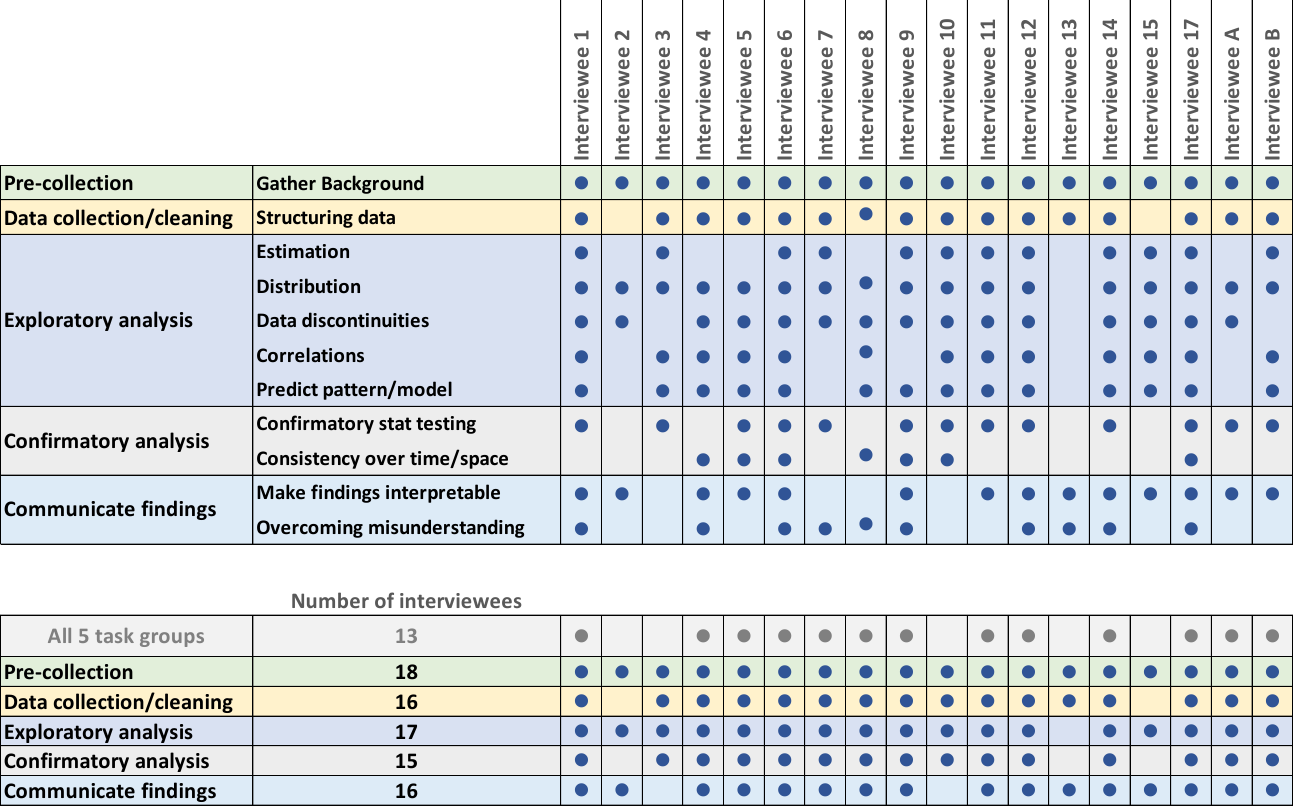}
    \caption{\textbf{Analytic tasks.}
    Participant coding for the Tasks perspective.
    }
    \label{tab:tasks}
\end{table}

\subsection{Analytical Tasks}

The Tasks perspective derived from the initial review of researcher notes proved to be well supported by subsequent formal coding processes.
Table~\ref{tab:tasks} gives an overview.
Thirteen of 18 participants reported performing work steps which fell within all 5 of the proposed tasks, and no participant reported fewer than 3.

Pre-collection activities, such as meeting with clients to determine their needs, gathering background information on available datasets, or proposing analytic methods, were universally reported among participants, with other steps nearly so.

Note that to qualify for the study, all participants began by confirming that they performed statistical testing as part of their regular work process, or had at some point.
Similarly, 17 of 18 participants reported publishing their work publicly, and the 18th reported sharing their work internally within their organization, all of which constitutes communicating results.
Therefore, it is likely that while these steps were not universally captured during the coding exercise, all participants did, in fact, perform these steps during their work.
This may further substantiate the Tasks perspective.

\begin{table}[htb]
    \centering
    \includegraphics[width=\linewidth]{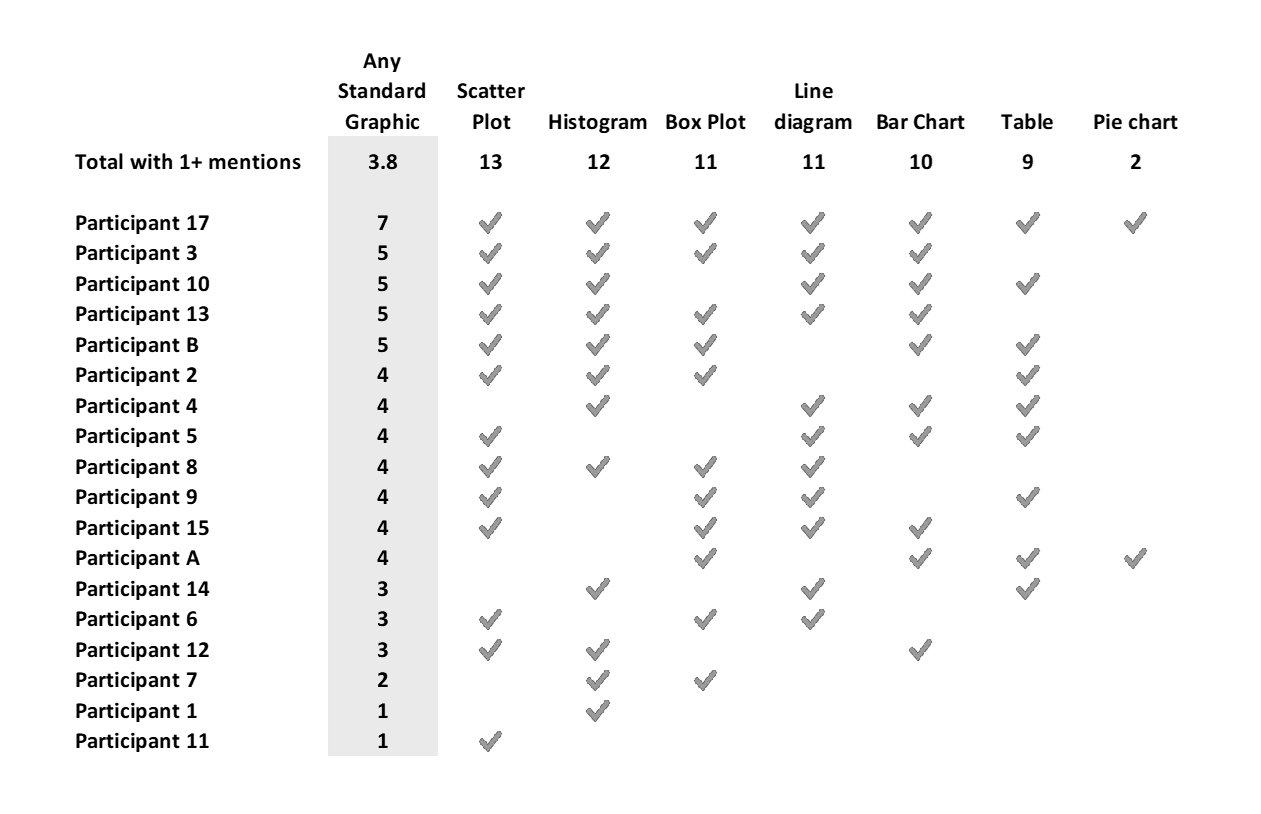}
    \caption{\textbf{Chart usage.}
    Incidences of chart usage per participant.
    }
    \label{tab:charts}
\end{table}

\subsection{Tools Reported}

Parsing transcripts for mentions of specific analytic tools (named methods, procedures, or statistical routines encapsulated within software packages) resulted in three lists: Broad-focus, narrow-focus, and symbolic tools.
Unsurprisingly given the visual-analytic focus of these conversations, visual tools outnumbered purely symbolic ones.
However, we were surprised by the diversity of narrow-focus tools.

Some broad-focus statistical graphics have a long history, wide availability in software packages, and ubiquitous appearances in literature with statistical content.
The authors interpret use of such tools as an indicator of the degree to which participants fold visualization into their work.
While we captured all mentions of visualization (see the supplemental material for details), we focused on seven graphic forms based on frequency and familiarity: \textsc{Scatterplot}, \textsc{Histogram}, \textsc{Boxplot}, \textsc{Line Diagram}, \textsc{Bar Chart}, \textsc{Table},\footnote{While numeric in content, tables make use of a visual schema~\cite{DBLP:journals/tvcg/BartramCT22}.} and \textsc{Pie Chart}.

Table~\ref{tab:charts} gives an overview of this list.
All 18 participants used some visualization from this list, likely with analytic intent.
Scatterplots were the most widely named graphic form, mentioned by 13 of 18 participants.
Histograms followed closely, with 12.
Both of these forms are explicitly analytic in function compared with other forms (pie charts) which are less useful for analysis but are sometimes favored in communicating findings.
Among participants reporting only 1 tool used, it was either a Scatterplot or Histogram.
All graphic forms on this list were widely used with the exception of the Pie Chart (2/18).
On average, participants reported use 3.8 out of these 7 tools. 

\begin{figure*}[htb]
    \centering
    \includegraphics[width=\linewidth]{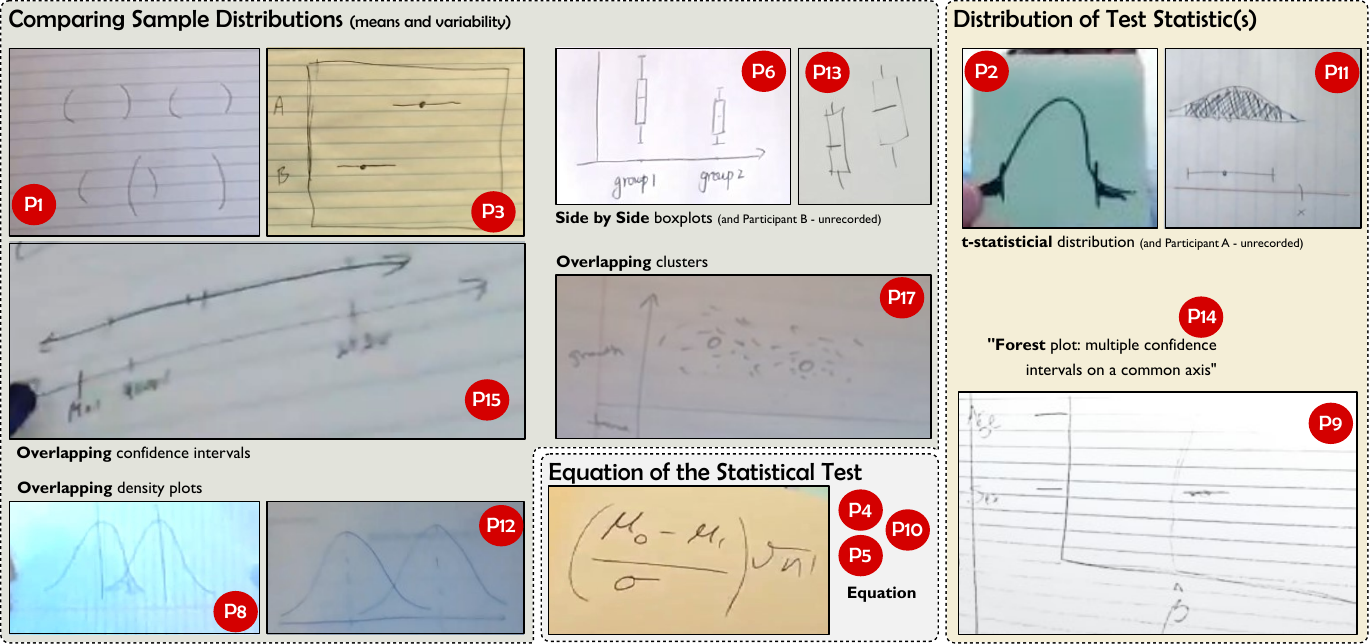}
    \caption{\textbf{Graphic elicitation examples.}
    Examples of graphics for statistical inference elicited from participants.}
    \label{fig:graphic-elicitation}
\end{figure*}

\subsection{Graphic Elicitation}

Twelve of 18 participants provided sketches representing what a statistical test looks like to them.  
Of the remaining six, two participants’ sketches (P-A and P-B) were captured via researcher notes.
Some participants indicated that they did not have a mental image other than the definitions/equations of the test.  
Thus, they had nothing to draw.

All participants who provided a sketch were asked to walk the researcher through their work.
These walkthroughs and sketches have been summarized in Figure~\ref{fig:graphic-elicitation}.
In total, 9 participants communicated that their vision of a statistical test was a pair of sample centers (mean or median) displayed side by side, with some indicator of sample variability around those centers (such as overlapping confidence intervals or side-by-side boxplots).
Five participants indicated they envisioned the distribution of the relevant test statistic, with the location of the realized test statistic noted, and with an indicator of the likelihood of being at that location on the distribution.
Two participants expressed that their mental image of a statistical test was that of an equation and provided no sketch (a third drew an equation).
One participant named a Forrest plot as their internal image of a statistical test, and provided an example from the literature.
One participant did not participate in the graphic elicitation portion of the interview.

%% file: content/05-discussion.tex
\section{Discussion}

Here we report on our findings from the interview study, including the role of visualization in statistics, the use need for evidentiary warrants, and our design guidelines.
We also discuss the limitations of our work.

\subsection{Visualization According to Statisticians}

\paragraph{Statisticians make extensive use of visualization.}

Every participant in this study employed visualization as a regular part of their analytic process.
All 18 made use of one or more of the common broad-use visualizations.
Most used several.
Most also used narrow-use visualizations; 14 of 18 participants reported making use of at least one specialized visualization with a narrow analytic focus.
Indeed, participants reporting using the fewest of the common broad-focus visualizations all reported making use of a specialized narrow-use visualization.
A third of participants (6/18) shared observations on the limits of data visualizations while still making use of them.

\paragraph{Statisticians often conceptualize their work in visual terms.}

Fourteen of 17 participants indicated that their internal conception of at least one inferential method is visual. 
Eleven of 16 participants felt that visualization \textit{is} analysis, 
fully integrated into their quantitative work.

\vspace{-3mm}
\begin{hquote}{yellow}
    \textit{Maybe I just take that for granted, the image thing.
    I think once we have data and the first thing probably is to plot the data and see what [it] looks like.
    I think [...] to understand the data, visualization is a very important tool.} (P6)
\end{hquote}
\vspace{-2mm}

\paragraph{Visualizations are not always shared.}

Yet despite their frequent use of visualization and their understanding of their analyses in visual terms, it is not necessarily the visualization work they share with others.  

\vspace{-3mm}
\begin{hquote}{yellow}
    \textit{Oh yes.
    I use line graphs.
    I use bar graphs.
    I use scatterplots.
    Absolutely.
    [...]
    Not something I would share typically...
    } (P13)
\end{hquote}
\vspace{-2mm}

\paragraph{Not good enough for visualization.}

Some participants described using visualization, but did not think that counted as visualization.
Two (P4 and P13) felt that artifacts which rose to the level of deserving the label, ``visualization,'' had to exceeded mere utilitarian analysis and achieve aesthetic value, something they had little confidence they could themselves create: ``\textit{I am really bad at visualizing things.}'' (P13)

\paragraph{Charts are for communication.}

Statisticians find visualization an indispensable tool for communicating results.
Ten analysts talked about visualization’s ability to communicate stories:  

\vspace{-3mm}
\begin{hquote}{yellow}
    \textit{You have your analysis, your P-value, your confidence intervals, your hypothesis test, what decision did you make, all of that.
    That's not what I'm going to put in a presentation typically. 
    It's going to be the graph. [...]
    It [numeric representation of results] is important, I'm not trying to minimize the importance of it, but it's not what people understand when you're trying to communicate a result [...] unless you're in a room with statisticians.} (P12)
\end{hquote}
\vspace{-2mm}

\paragraph{Chart or calculation?}

For some, visualization may be preferable to purely calculative methods.
Five participants reported that visualization provides more powerful evidence than numeric methods:

\vspace{-3mm}
\begin{hquote}{yellow}
    \textit{I tell every class I teach when we go through inference, do the graphs first.
    And if the inference that you do doesn't support what you saw in the graph, something's terribly wrong.} (P12)
\end{hquote}
\vspace{-2mm}

Yet nearly as many (4) reported the opposite, that math is a greater evidentiary warrant, and gave a reason, namely, that visualization relies too much upon the judgement of the analyst. 
As participant 5 said, ``\textit{Yeah, you don't want to rely on your eyes.
Having a number is better.}''

Other strikes against visualization include the primacy of calculative methods in achieving publication.
Two participants reported some version of being wary of aspects of p-value based statistical tests, but using them because publication required it.

\paragraph{The cost of vis.}

Visualization can also result in a lot of work for the statistician, without necessarily earning them concomitant rewards.
Intuitive design is critical for visualizations meant to communicate findings, but intuitive design tends to disappear from view.
Thus, a statistician who works hours creating a display which their client can understand with a glance is, in effect, hiding their work.
This is the ``Designer’s dilemma,'' where the more successful the work, the less it might be noticed (or appreciated) by its consumers:

\vspace{-3mm}
\begin{hquote}{yellow}
    \textit{So, [clients] think maybe you hand them a graph and the analysis results and [they think], ``Oh, that probably took them 20 minutes to do that.''
    [...] But they miss all the hours and hours of work that go into that final result.} (P12)
\end{hquote}

\paragraph{A strained relationship.}

The relationship between statisticians and visualization is complex, despite visualization apparently being integrated into every phase of a statistician’s workflow.
Different practitioners give more or less emphasis to its use, though all the participants in this study report using visualization for at least some phases of their work.
While it is possible that this outcome is a result of the Hawthorne effect, we note that our interview protocol used deliberately open-ended questions. 
Our participants were also seasoned professionals and not impressionable novices.
They variously reported using process diagrams for planning analyses, visual exploratory data analysis for both data cleaning and hypothesis formation, specialized narrow-use visualizations during confirmatory analysis either as a reasonableness check or a source of primary findings, and, finally, communication of results.
Further, most think about their analyses in visual terms, and find in visualization a powerful tool for supporting evidence-based results.  

P3 may provide the clearest example of the conflicted relationship statisticians appear to have with visualization. 
This person reported both a preference for calculative math over visualization, and vice versa.
On the one hand, in referring to one of the strawmen graphics:

\vspace{-3mm}
\begin{hquote}{yellow}
    \textit{So it seems like it's more subjective in a way.
    And I don't think that a new statistician or anyone who's been through the school that I went through [...] has seen this often enough to make the best decisions with it.} (P3)
\end{hquote}
\vspace{-3mm}

But on the other hand, when describing the process of understanding a hypothetical regression output during their workflow:

\vspace{-3mm}
\begin{hquote}{yellow}
    \textit{Then I see where's the plot?
    How does it look here?
    What's the shape?
    So all those things go on in my head.} (P3)
\end{hquote}

\subsection{Evidentiary Warrants}

\paragraph{The human element.}

If visualization acts as interface between the human and computer in extended cognition~\cite{DBLP:journals/ijmms/ScaifeR96}, then the resulting compound system hinges upon human judgment---our visual intuitions, visual acuity, subject matter pre-knowledge, and imagination.
It is thus subject to human biases, mental blind spots, and the various shortcomings of our eyes, as pointed out by 5/16 participants (4 describing visualization as too subjective, and a fifth making much the same point when suggesting reliance upon visual systems can reduce validity).  

By contrast, traditional, equation-based statistical tests fully externalize the critical decision point, ostensibly removing the human element.
Provided results criteria are selected in advance (said 6/16 participants), and all test assumptions are met (3/16), results from statistical tests \textit{feel} more objective.
Five of 16 participants described relying upon statistical tests to provide evidentiary warrants for their findings, despite nearly as many warning that p values encourage bad thinking (4/16).

For example, participant 15 expressed both ideas.
On the one hand, statistical tests encourage dichotomous thinking about situations:

\vspace{-3mm}
\begin{hquote}{yellow}
    \textit{There's nothing wrong with frequentist methods [such as parametric statistical tests], but it's kind of black and white.} (P15)
\vspace{-2mm}
\end{hquote}

Yet at the same time:

\vspace{-3mm}
\begin{hquote}{yellow}
    \textit{... [valid statistical testing] tells me whether my results should be actionable and meaningful to someone... the actual [sample] estimates themselves being different does not necessarily mean that things truly are different in the population.} (P15)
\end{hquote}
\vspace{-2mm}

This reliance upon calculative statistical tests and distrust of visualization appears to conflict with visualization’s wide use as an analytic method by participants; it is in direct conflict with participants expressions that visualization is an important check on the tests themselves.  

\paragraph{A resolution of apparent conflicts.}

A central thread may explain this conflict: the statisticians in our study argue against their own judgements wherever they can, constantly seeking to verify their findings with multiple independent methods.
In short, it is not that these statisticians trust this or that method and distrust another---they distrust them all, or more precisely, never trust any one method by itself.

10 of 16 participants expressed the idea that they should always distrust their own findings.
For example:

\vspace{-2mm}
\begin{hquote}{yellow}
    \textit{So, you want to use simulation to test that and also compare to the previous methods [...]
    and then you apply your methods [...] 
    and at this time you also need to talk with your collaborators [...] 
    And then they will help you to evaluate whether the results make sense or not. [...]  
    And then you want to analyze whether it is because there is a bug in your method or in your code [...]
    } (P6)
\end{hquote}
\vspace{-2mm}

\vspace{-2mm}
\begin{hquote}{yellow}
    \textit{There's always a reason. [...]
    It sometimes will come down to sample size or the overall variation if it's a two-sample test of maybe one group has larger variation than the other.
    [...]
    Maybe I'm wrong, but in my head there's always a reason...} (P12)
\end{hquote}

When numbers would seem to support their favored hypotheses, these statisticians look to visualization to see whether unknown outliers, gaps, or the like, explain away their findings.
But if they see a pattern in a visualization, they look to calculative methods to act as a check on their eyes.  
Presented with an apparent difference between sample means, they test whether there is any likelihood that random chance can explain that away.
They check their assumptions (7/16 participants) and conduct parallel analyses (4/16).
With already checked and tested findings in hand, they ask subject matter experts whether the results make sense to them (3/16 participants). 

In each of these cases, we speculate that the statisticians are seeking ways to undercut apparent findings.
They pit one analytic method against another, and all must line up for the experienced analyst to accept a finding as probably, or even possibly, true.    

\paragraph{Staying connected to reality.}

Among the most frequently expressed understanding captured during this study is the constant effort by participants to link their results back to the reality they are meant to describe.
Fourteen of 16 participants discussed this during their interviews.
For example, said P9: ``\textit{if necessary, go back and revise the statistical analysis plan in light of the reality of the situation as we've discovered it.}'' 
P11 put it this way: ``\textit{... you should trust the data and not come in with [...] strong priors.}''
P17's thoughts were shared by many: ``\textit{And I find generally data will tell the truth.}''

\paragraph{Effect sizes over math.}

Five participants (p1, p5, p6, p8, p13) talked about the validity of their results hinging upon a clear and well understood link between their measures and their phenomena of interest. 
Indeed, seven participants described how thinking through facts was key to understanding the math of their analyses.
For example:

\vspace{-3mm}
\begin{hquote}{yellow}
    \textit{
    I don't understand statistics.
    I will freely admit I don't understand it, purely mathematically.
    I understand it if I'm looking at something tangible that has meaning to me...} (P1)
\end{hquote}
\vspace{-3mm}

It appears that a key outgrowth of this reality-focus is a preference among many statisticians for privileging effect sizes over p-values.
Nine of 16 participants expressed this idea.
Effect sizes are typically some ratio of the difference in a key measure to the variability in that measure, where large values indicate analytically important, rather than merely statistically improbable, results.
Effect sizes thus concern the practical significance of a finding, rather than statistical (or probabilistic) significance.
In the current climate of concern over p-value statistical testing, focusing on effect sizes is one potential answer.
This is also what current best practices in statistical reporting suggests~\cite{Calin-Jageman2019}.
It is also supported by prior work showing that confidence intervals can cause people to overestimate effect sizes~\cite{DBLP:conf/chi/HofmanGH20}, and that more sophisticated visual representations beyond typical error bars are needed to convey effect size nuances~\cite{DBLP:journals/tvcg/CorrellG14}.
If nothing else, focusing on effect sizes necessarily means focusing on the reality of the subject. 

\paragraph{Uncertainty is certain.}

Nine of the participants expressed their understanding of a t-test as some version of a comparison of overlapping confidence intervals, i.e., sample means with some indicator of variability.
This approach emphasizes the samples themselves, rather than the unseen population they are meant to represent.

The t-test and p-value  focus ultimately on the population, by posing the likelihood of drawing such a sample if the population were random with regard to some independent variable.
Yet the majority of participants in this study keep their focus on the samples, while asking whether their results achieve statistical significance, and thus provide an evidentiary warrant to make statements about the population.  
The practical result is logically equivalent, but speaks to the difficulty of statistical inference.
In many cases, even the experts don't fully embrace the meaning of the mathematics they rely upon, instead mentally falling back on simpler heuristics. 


\subsection{Design Recommendations}

Given statisticians' own use of visualization in their analytic process, it appears that, as a general approach, visualizing inferential statistics is acceptable to the statistical community.
However, it also appears that no single visualization approach will have the confidence of that community.
Rather, visualization should be paired with other confirmatory tools.
This will provide the analyst both an intuitive understanding of the data (visual), and ``more objective'' numbers (where the decision point is determined by the calculation rather than the user's eye).

Furthermore, our findings indicate that visual inferential tools should include an indicator of effect size that is declared in advance of seeing the data, just like alpha (minimum acceptable p-value), to avoid post-hoc rationalization.
Selecting effect sizes requires users to understand their data, usually by speaking to subject matter experts.
It also automatically combats dichotomous thinking, as having two measures to choose (alpha and effect size) turns the significance decision multi-dimensional.
We base this recommendation on prior work showing that visualizations that show nuanced aspects of data can reduce dichotomous thinking~\cite{DBLP:journals/tvcg/HelskeHCYB21}, as well as general inferential statistics~\cite{Sullivan2012}.

\subsection{Limitations}

Our study in this paper involved a total of 18 professional statisticians, but while we took care to choose participants from many different fields, educational backgrounds, and demographics, there are certainly several threats to generalizing these results too widely. 
Given the qualitative and highly personal nature of these practices, we feel that it can be hard to draw conclusive findings from our work. 
Furthermore, as discussed in our positionality statement (Section~\ref{sec:positionality}), we ourselves as researchers represent only a small fraction of the worldwide statistician population.

Our study was focused on parametric inferential statistics.
Statistical inference is obviously a much larger field, and includes topics such as non-parametric and Bayesian methods.
This may limit the generality of our design recommendations and suggests avenues for future research.

While design probes have been proven effective because they provide a common ground for discussion~\cite{DBLP:conf/chi/WallaceMWO13}, which can be helpful for laypersons, they may also constrain ideation.
Furthermore, our strawman graphics (Figure~\ref{fig:strawman-graphics}) are not radical or even particularly novel, and perhaps a more radical set of design ideas could have sparked more innovation.
However, we took care to ask for graphic elicitation (Phase II) prior to showing our graphics (Phase III) to avoid biasing the participants.
Furthermore, the purpose of this study was mostly to understand the mindsets and practices of professional statisticians, and more effort will be needed to develop these graphics in the future.

%% file: content/06-conclusion.tex
\section{Conclusion}

We have presented results from a qualitative interview study involving 18 professional participants with the goal of understanding their use of visualization (RQ1), mental models of inferential statistics (RQ2), and thoughts on designs for visual inference (RQ3). 
Our findings, which were coded and summarized from interview transcripts, suggest a significant influence of visualization even in the workflows of statisticians who self-report as ``traditionalists.''
In fact, many of their mental models of statistical inference appear to be at least somewhat visually based.
We use these findings to suggest several design guidelines for how to design new statistical and visualization tools that can help people make sense of their own data.

%% file: vis-stat.bbl
\begin{thebibliography}{32}
\providecommand{\natexlab}[1]{#1}
\providecommand{\url}[1]{\texttt{#1}}
\expandafter\ifx\csname urlstyle\endcsname\relax
  \providecommand{\doi}[1]{doi: #1}\else
  \providecommand{\doi}{doi: \begingroup \urlstyle{rm}\Url}\fi

\bibitem[Albers et~al.(2014)Albers, Correll, and Gleicher]{Albers2014}
Danielle Albers, Michael Correll, and Michael Gleicher.
\newblock Task-driven evaluation of aggregation in time series visualization.
\newblock In \emph{Proceedings of the {ACM} Conference on Human Factors in
  Computing Systems}, pages 551--560, New York, NY, USA, 2014. {ACM}.
\newblock \doi{10.1145/2556288.2557200}.
\newblock URL \url{https://doi.org/10.1145/2556288.2557200}.

\bibitem[Bartram et~al.(2022)Bartram, Correll, and
  Tory]{DBLP:journals/tvcg/BartramCT22}
Lyn Bartram, Michael Correll, and Melanie Tory.
\newblock Untidy data: The unreasonable effectiveness of tables.
\newblock \emph{{{IEEE} Transactions on Visualization and Computer Graphics}},
  28\penalty0 (1):\penalty0 686--696, 2022.
\newblock \doi{10.1109/TVCG.2021.3114830}.

\bibitem[Beecham et~al.(2017)Beecham, Dykes, Meulemans, Slingsby, Turkay, and
  Wood]{Beecham2017}
Roger Beecham, Jason Dykes, Wouter Meulemans, Aidan Slingsby, Cagatay Turkay,
  and Jo~Wood.
\newblock Map lineups: Effects of spatial structure on graphical inference.
\newblock \emph{{{IEEE} Transactions on Visualization and Computer Graphics}},
  23\penalty0 (1):\penalty0 391--400, 2017.
\newblock \doi{10.1109/TVCG.2016.2598862}.
\newblock URL \url{https://doi.org/10.1109/TVCG.2016.2598862}.

\bibitem[Buja et~al.(2009)Buja, Cook, Hofmann, Lawrence, Lee, Swayne, and
  Wickham]{Buja2009}
Andreas Buja, Dianne Cook, Heike Hofmann, Michael Lawrence, Eun-Kyung Lee,
  Deborah~F. Swayne, and Hadley Wickham.
\newblock Statistical inference for exploratory data analysis and model
  diagnostics.
\newblock \emph{Philosophical Transactions of the Royal Society}, 367\penalty0
  (1906):\penalty0 4361--4383, 2009.
\newblock \doi{10.1098/rsta.2009.0120}.
\newblock URL \url{https://doi.org/10.1098/rsta.2009.0120}.

\bibitem[Calin-Jageman and Cumming(2019)]{Calin-Jageman2019}
Robert~J. Calin-Jageman and Geoff Cumming.
\newblock The new statistics for better science: Ask how much, how uncertain,
  and what else is known.
\newblock \emph{The American Statistician}, 73\penalty0 (sup1):\penalty0
  271--280, 2019.
\newblock \doi{10.1080/00031305.2018.1518266}.

\bibitem[Casella and Berger(2001)]{casella2001statistical}
George Casella and Roger~L Berger.
\newblock \emph{Statistical Inference}.
\newblock Cengage Learning, Belmont, CA, USA, 2nd edition, 2001.

\bibitem[Cleveland(1993)]{Cleveland1993}
William~S. Cleveland.
\newblock \emph{Visualizing Data}.
\newblock Hobart Press, Summit, NJ, USA, 1993.

\bibitem[Correll and Gleicher(2014)]{DBLP:journals/tvcg/CorrellG14}
Michael Correll and Michael Gleicher.
\newblock Error bars considered harmful: Exploring alternate encodings for mean
  and error.
\newblock \emph{{{IEEE} Transactions on Visualization and Computer Graphics}},
  20\penalty0 (12):\penalty0 2142--2151, 2014.
\newblock \doi{10.1109/TVCG.2014.2346298}.

\bibitem[Correll and Heer(2017)]{conf/chi/CorrellH17}
Michael Correll and Jeffrey Heer.
\newblock Regression by eye: Estimating trends in bivariate visualizations.
\newblock In \emph{Proceedings of the {ACM} Conference on Human Factors in
  Computing Systems}, pages 1387--1396, New York, NY, USA, 2017. {ACM}.
\newblock \doi{10.1145/3025453.3025922}.
\newblock URL \url{https://doi.org/10.1145/3025453.3025922}.

\bibitem[Correll et~al.(2012)Correll, Albers, Franconeri, and
  Gleicher]{Correll2012}
Michael Correll, Danielle Albers, Steven Franconeri, and Michael Gleicher.
\newblock Comparing averages in time series data.
\newblock In \emph{Proceedings of the {ACM} Conference on Human Factors in
  Computing Systems}, pages 1095--1104, New York, NY, USA, 2012. {ACM}.
\newblock \doi{10.1145/2207676.2208556}.
\newblock URL \url{https://doi.org/10.1145/2207676.2208556}.

\bibitem[Correll et~al.(2019)Correll, Li, Kindlmann, and
  Scheidegger]{Correll2019}
Michael Correll, Mingwei Li, Gordon~L. Kindlmann, and Carlos Scheidegger.
\newblock Looks good to me: Visualizations as sanity checks.
\newblock \emph{{{IEEE} Transactions on Visualization and Computer Graphics}},
  25\penalty0 (1):\penalty0 830--839, 2019.
\newblock \doi{10.1109/TVCG.2018.2864907}.
\newblock URL \url{https://doi.org/10.1109/TVCG.2018.2864907}.

\bibitem[Fisher(1922)]{Fisher1922}
Ronald Fisher.
\newblock On the mathematical foundations of theoretical statistics.
\newblock \emph{Philosophical Transactions of the Royal Society of London},
  222\penalty0 (594--604), 1922.
\newblock \doi{https://doi.org/10.1098/rsta.1922.0009}.

\bibitem[Gaver et~al.(1999)Gaver, Dunne, and
  Pacenti]{DBLP:journals/interactions/GaverDP99}
William~W. Gaver, Anthony Dunne, and Elena Pacenti.
\newblock Design: Cultural probes.
\newblock \emph{Interactions}, 6\penalty0 (1):\penalty0 21--29, 1999.
\newblock \doi{10.1145/291224.291235}.

\bibitem[Gleicher et~al.(2013)Gleicher, Correll, Nothelfer, and
  Franconeri]{Gleicher2013}
Michael Gleicher, Michael Correll, Christine Nothelfer, and Steven Franconeri.
\newblock Perception of average value in multiclass scatterplots.
\newblock \emph{{{IEEE} Transactions on Visualization and Computer Graphics}},
  19\penalty0 (12):\penalty0 2316--2325, 2013.
\newblock \doi{10.1109/TVCG.2013.183}.
\newblock URL \url{https://doi.org/10.1109/TVCG.2013.183}.

\bibitem[Grammel et~al.(2010)Grammel, Tory, and
  Storey]{DBLP:journals/tvcg/GrammelTS10}
Lars Grammel, Melanie Tory, and Margaret{-}Anne~D. Storey.
\newblock How information visualization novices construct visualizations.
\newblock \emph{{{IEEE} Transactions on Visualization and Computer Graphics}},
  16\penalty0 (6):\penalty0 943--952, 2010.
\newblock \doi{10.1109/TVCG.2010.164}.

\bibitem[Hald(1998)]{Hald1998}
Anders Hald.
\newblock \emph{A History of Mathematical Statistics from 1750 to 1930}.
\newblock Probability and Statistics. Wiley, 1998.

\bibitem[Helske et~al.(2021)Helske, Helske, Cooper, Ynnerman, and
  Besan{\c{c}}on]{DBLP:journals/tvcg/HelskeHCYB21}
Jouni Helske, Satu Helske, Matthew Cooper, Anders Ynnerman, and Lonni
  Besan{\c{c}}on.
\newblock Can visualization alleviate dichotomous thinking? effects of visual
  representations on the cliff effect.
\newblock \emph{{{IEEE} Transactions on Visualization and Computer Graphics}},
  27\penalty0 (8):\penalty0 3397--3409, 2021.
\newblock \doi{10.1109/TVCG.2021.3073466}.

\bibitem[Hofman et~al.(2020)Hofman, Goldstein, and
  Hullman]{DBLP:conf/chi/HofmanGH20}
Jake~M. Hofman, Daniel~G. Goldstein, and Jessica Hullman.
\newblock How visualizing inferential uncertainty can mislead readers about
  treatment effects in scientific results.
\newblock In \emph{Proceedings of the {ACM} Conference on Human Factors in
  Computing Systems}, pages 1--12, New York, NY, USA, 2020. {ACM}.
\newblock \doi{10.1145/3313831.3376454}.

\bibitem[Huron et~al.(2014)Huron, Jansen, and
  Carpendale]{DBLP:journals/tvcg/HuronJC14}
Samuel Huron, Yvonne Jansen, and Sheelagh Carpendale.
\newblock Constructing visual representations: Investigating the use of
  tangible tokens.
\newblock \emph{{{IEEE} Transactions on Visualization and Computer Graphics}},
  20\penalty0 (12):\penalty0 2102--2111, 2014.
\newblock \doi{10.1109/TVCG.2014.2346292}.

\bibitem[Lazar et~al.(2017)Lazar, Feng, and Hochheiser]{DBLP:books/el/LFH2017}
Jonathan Lazar, Jinjuan Feng, and Harry Hochheiser.
\newblock \emph{Research Methods in Human-Computer Interaction}.
\newblock Morgan Kaufmann, San Francisco, CA, USA, 2nd edition, 2017.

\bibitem[Lehmann and Romano(2005)]{Lehmann2005}
Eric~L. Lehmann and Joseph~P. Romano.
\newblock \emph{Testing Statistical Hypotheses}.
\newblock Springer, New York, NY, USA, 3rd edition, 2005.

\bibitem[Mustafa(1996)]{Mustafa1996}
R.~Yilmaz Mustafa.
\newblock The challenge of teaching statistics to non-specialists.
\newblock \emph{Journal of Statistics Education}, 4\penalty0 (1), 1996.
\newblock \doi{10.1080/10691898.1996.11910504}.

\bibitem[Newburger et~al.(2023)Newburger, Correll, and Elmqvist]{Newburger2023}
Eric Newburger, Michael Correll, and Niklas Elmqvist.
\newblock Fitting bell curves to data distributions using visualization.
\newblock \emph{{{IEEE} Transactions on Visualization and Computer Graphics}},
  2023.
\newblock \doi{10.1109/TVCG.2022.3210763}.

\bibitem[Pousman et~al.(2007)Pousman, Stasko, and
  Mateas]{DBLP:journals/tvcg/PousmanSM07}
Zachary Pousman, John~T. Stasko, and Michael Mateas.
\newblock Casual information visualization: Depictions of data in everyday
  life.
\newblock \emph{{{IEEE} Transactions on Visualization and Computer Graphics}},
  13\penalty0 (6):\penalty0 1145--1152, 2007.
\newblock \doi{10.1109/TVCG.2007.70541}.

\bibitem[Scaife and Rogers(1996)]{DBLP:journals/ijmms/ScaifeR96}
Michael Scaife and Yvonne Rogers.
\newblock External cognition: how do graphical representations work?
\newblock \emph{International Journal of Human-Computer Studies}, 45\penalty0
  (2):\penalty0 185--213, 1996.
\newblock \doi{10.1006/ijhc.1996.0048}.

\bibitem[Schervish(1995)]{Schervish1995}
Mark~J. Schervish.
\newblock \emph{Theory of Statistics}.
\newblock Springer Verlag, New York, NY, USA, 1995.
\newblock \doi{10.1007/978-1-4612-4250-5}.

\bibitem[Shneiderman(1996)]{DBLP:conf/vl/Shneiderman96}
Ben Shneiderman.
\newblock The eyes have it: {A} task by data type taxonomy for information
  visualizations.
\newblock In \emph{Proceedings of the {IEEE} Symposium on Visual Languages},
  pages 336--343, Los Alamitos, CA, USA, 1996. {IEEE Computer Society}.
\newblock \doi{10.1109/VL.1996.545307}.

\bibitem[Student(1908)]{Student1908}
Student.
\newblock The probable error of a mean.
\newblock \emph{Biometrika}, 6\penalty0 (1):\penalty0 1--–25, 1908.
\newblock \doi{10.1093/biomet/6.1.1}.
\newblock URL \url{https://doi.org/10.1093/biomet/6.1.1}.

\bibitem[Sullivan and Feinn(2012)]{Sullivan2012}
Gail~M. Sullivan and Richard Feinn.
\newblock Using effect size---or why the {P} value is not enough.
\newblock \emph{Journal of Graduate Medical Education}, 4\penalty0
  (3):\penalty0 279--282, 2012.
\newblock \doi{10.4300/JGME-D-12-00156.1}.

\bibitem[Tukey(1977)]{Tukey1977}
John~W. Tukey.
\newblock \emph{Exploratory Data Analysis}.
\newblock Addison-Wesley, Reading, MA, USA, 1977.

\bibitem[Wallace et~al.(2013)Wallace, McCarthy, Wright, and
  Olivier]{DBLP:conf/chi/WallaceMWO13}
Jayne Wallace, John~C. McCarthy, Peter~C. Wright, and Patrick Olivier.
\newblock Making design probes work.
\newblock In \emph{Proceedings of the {ACM} Conference on Human Factors in
  Computing Systems}, pages 3441--3450, New York, NY, USA, 2013. {ACM}.
\newblock \doi{10.1145/2470654.2466473}.

\bibitem[Wickham et~al.(2010)Wickham, Cook, Hofmann, and Buja]{Wickham2010}
Hadley Wickham, Dianne Cook, Heike Hofmann, and Andreas Buja.
\newblock Graphical inference for infovis.
\newblock \emph{{{IEEE} Transactions on Visualization and Computer Graphics}},
  16\penalty0 (6):\penalty0 973--979, 2010.
\newblock \doi{10.1109/TVCG.2010.161}.
\newblock URL \url{https://doi.org/10.1109/TVCG.2010.161}.

\end{thebibliography}
